\documentclass[10pt]{article}

\usepackage{amsmath}
\usepackage{amssymb}

\usepackage{graphicx}

\usepackage{cite}

\usepackage{color} 

\usepackage{setspace} 
\usepackage[labelfont=bf,labelsep=period,justification=raggedright]{caption}

\makeatletter
\renewcommand{\@biblabel}[1]{\quad#1.}
\makeatother

\date{}

\begin{document}

\begin{flushleft}
{\Large
\textbf{Slow sedimentation and deformability of charged lipid vesicles}
}
\\
Iv\'an Rey Su\'arez$^{1}$, 
Chad Leidy$^{2}$, 
Gabriel T\'ellez$^{2}$,
Guillaume Gay$^{3}$,
Andres Gonzalez-Mancera$^{1,\ast}$
\\
\bf{1} Department of Mechanical Engineering, Universidad de los Andes, Bogota, Colombia.
\\
\bf{2} Department of Physics, Universidad de los Andes, Bogota, Colombia.
\\
\bf{3} Universite Paul Sabatier Toulose III, Toulose - France
\\
$\ast$ E-mail: Corresponding angonzal@uniandes.edu.co
\end{flushleft}

\section*{Abstract}

The study of vesicles in suspension is important to understand the complicated dynamics exhibited by cells in \emph{in vivo} and \emph{in vitro}. We developed a computer simulation based on the boundary-integral method to model the three dimensional gravity-driven sedimentation of charged vesicles towards a flat surface. The membrane mechanical behavior was modeled using the Helfrich Hamiltonian and near incompressibility of the membrane was enforced via a model which accounts for the thermal fluctuations of the membrane. The simulations were verified and compared to experimental data obtained using suspended vesicles labelled with a fluorescent probe, which allows visualization using fluorescence microscopy and confers the membrane with a negative surface charge. The electrostatic interaction between the vesicle and the surface was modeled using the linear Derjaguin approximation for a low ionic concentration solution. The sedimentation rate as a function of the distance of the vesicle to the surface was determined both experimentally and from the computer simulations. The gap between the vesicle and the surface, as well as the shape of the vesicle at equilibrium were also studied. It was determined that inclusion of the electrostatic interaction is fundamental to accurately predict the sedimentation rate as the vesicle approaches the surface and the size of the gap at equilibrium, we also observed that the presence of charge in the membrane increases its rigidity.

\section*{Introduction}

Lipid vesicles in the same size range as cells are widely used to investigate the physical properties of cell membranes, providing a simplified model system to investigate the role that the membrane plays in the mechanical deformation of cells. It is important to consider lipid membrane mechanics in order to understand many relevant processes involving cell deformation. For example, red blood cells (RBCs) undergo large deformations in the microcirculation~\cite{tomaiuolo:2011us} which appears to optimize gas exchange between blood and the surrounding tissues~\cite{de2007plasma,yoon2009flickering}. Given that unilamellar lipid vesicles share some features with RBCs, such as the ability to generate equilibrium biconcave shapes and dynamics under shear flow such as tank-threading and tumbling, the study of its behavior under flow conditions can provide useful information in the understanding of blood rheology~\cite{vlahovska:2009fe,kaoui:2009sx,kaoui:2009rm}.

Given the high stretching elastic modulus of lipid membranes, these are generally assumed to be incompressible (constant area per lipid molecule)~\cite{helfrich:1973eg,seifert:1995ui,seifert:1997zc}. As a consequence, the dynamics of vesicles under flow are sensitive to the value of the reduced volume, $\tau=6\sqrt{\pi}V/S^{3/2}$ ($V$ is the volume of the vesicle and $S$ its surface area)~\cite{biben:2011vy}, which quantifies the amount of area available for the deformation of the vesicle. The maximum value $\tau=1$ corresponds to a sphere and values of $\tau<1$ correspond to shapes with an initial excess of area (as compared to a sphere).

In order to enforce local incompressibility a local isotropic tension, in the form of a Lagrange multiplier, can be added to the force density acting on the membrane~\cite{sukumaran:2001py, misbah:2006jl, kaoui:2008ph, kaoui2011two, boedec20113d, zhao2011dynamics}. The value of the Lagrange multiplier can be calculated approximately by considering a nearly incompressible elastic membrane and calculating the tension due to deformation, or by enforcing that the resulting flow field has zero divergence on the surface of the membrane. Either method neglects the effect of thermal fluctuations which is generally justified due to the relatively high stresses the hydrodynamics impose on the deforming vesicles.

Lipid membrane deformability influences the movement of vesicles close to a surface~\cite{abkarian:2005ib}. For example, it was shown that deflated vesicles under shear flow at a critical shear rate are lifted from the surface by means of a force of viscous origin. After lifting from the surface, the vesicles acquire the shape of prolate ellipsoids as they move in the direction of the flow. A direct numerical simulation (DNS) using a spectral boundary integral equation method was used to quantitatively determine the lift force acting on deflated vesicles near a wall \cite{zhao2011dynamics}. It was demonstrated that lift velocity strongly depends on the vesicle's reduced volume and the viscosity ratio.

Despite the fact that the dynamic behavior of vesicles under external flow fields has been widely studied experimentally~\cite{kantsler:2005zo, kantsler:2006lq, deschamps:2009tt}, computationally~\cite{kaoui:2008ph, zhao2011dynamics,veerapaneni:2011aa}, and theoretically~\cite{misbah:2006jl, danker:2009ep}, sedimentation of lipid vesicles had received little attention. Recently, Boedec and coworkers studied the sedimentation of vesicles using computational~\cite{boedec20113d} and theoretical~\cite{boedec2012settling} methods. In both cases the contribution of thermal fluctuations is neglected and sedimentation is studied in an infinite space. Experimental observations of pear-like shapes and microtether extrusion of sedimenting vesicles have also been reported~\cite{huang2011sedimentation}.

Buoyancy-driven motion of drops has been widely studied in the fluid mechanics literature both experimentally and numerically. For example, the buoyancy-driven interaction of viscous drops was studied experimentally and numerically using the boundary integral method~\cite{kushner:2001pm}. The effect of surfactants on buoyancy-driven motion and interaction of viscous deformable drops was also studied using the boundary-integral method~\cite{rother:2004ll, rother:2006yk, rother:2008yv, ascoli:1990xy}. In all these works, the computational method has proven to be precise and in good agreement with both theoretical predictions and experimental data.

In this work we focus on the gravity induced sedimentation and deformation of initially quasi-spherical vesicles as they approach a flat horizontal no-slip surface. For this purpose we develop a computational algorithm based on the boundary integral method~\cite{pozrikidis:1992rz}. Our algorithm is mostly based on the method described by Zinchenko and coworkers~\cite{zinchenko:1997lk, zinchenko:1999pr} which was developed to study the interaction of deformable drops. More recently a similar algorithm has been used to study the dynamics of a vesicle in shear flow~\cite{biben:2011vy} and in a wall-bound shear flow \cite{zhao2011dynamics}.

The assumption of membrane incompressibility is based on the fact that direct area expansion is expensive from an energy point of view. Studies carried out by micropipette aspiration have shown that vesicles can undergo strain at low tensions. For example 1-stearoyl-2-oleoyl-phosphatidylcholine (SOPC) vesicles have shown strains of up to 0.06 for applied tensions of 0.5 mN/m~\cite{ly:2004fh}, this strain is accounted for by considering the smoothing of suboptical thermal fluctuations of the membrane in the low tension regime~\cite{evans:1990gl}. Therefore, even though changes in the lipid surface density are not expected to be significant, small membrane strain is likely as a result of smoothing of thermal undulations, which are accessible under low tension conditions. In other words, the thermal undulations act as an area reservoir which allows the membrane to deform under moderate forces.

In our model, membrane mechanics are dictated by the traditional Helfrich Hamiltonian~\cite{zhong-can:1989zn} and we take into account the smoothing of thermal undulations by relating area strain to tension by the constitutive function proposed by Evans and Rawicz~\cite{evans:1990gl}. This equation considers area strain to be caused by the superposition of the smoothing of thermal undulations \emph{plus} a direct expansion of the area per molecule. The latter requires much higher energy and is not available at low tensions, such as those exerted by gravity in our experiments.

Many experiments with lipid vesicles are performed in glucose/sucrose solutions on glass substrates, using fluorescent dyes to label the membrane. These studies do not control for ionic conditions, and usually ignore the influence of electrostatic interactions between different components in their experimental setups. These electrostatic interactions may affect the behavior of a suspended vesicle, for instance in determining the sedimentation rate close to the wall and the gap between the vesicle and the glass surface at equilibrium. In our work electrostatic interaction between the membrane and the planar surface are taken into account by using the Derjaguin, Landay, Verwey, Overbeek (DLVO) theory of colloidal stability~\cite{derjaguin:1940mk, carnie1994electrical, stankovich1996electrical}, where the linear Derjaguin solution for the Poisson-Boltzmann equation of the system was used to find the interaction force between the vesicle and the substrate.

By using single plane imaging microscopy (SPIM) we study the sedimentation dynamics of vesicles, this technique allows to follow a vesicle as a function of time with a better time resolution than a scanning technique such as confocal microscopy. We study the shapes and the size of the gap between the vesicle and the surface at equilibrium using confocal microscopy.

The objective of this work is to develop a robust computational model, which is able to accurately predict the behavior of charged vesicles sedimenting towards a charged flat surface. The model should contain the basic physics in order to be able to predict variables such as sedimentation rate as the vesicle approaches the surface and the gap between the vesicle and the glass surface at equilibrium.

\section*{Materials and methods} 
\label{sec:materials_and_methods}

\subsection*{Vesicle preparation and experimental setup} 
\label{sub:vesicle_preparation_and_experimental_setup}

1-Palmitoyl 2-oleoyl-phosphatidylcholine (POPC) was purchased from Avanti Polar Lipids (Alabaster, AL). \emph{N}-(7-nitrobenz-2-oxa-1, 3 diazol-4-yl)-1, 2 dihexadecanoyl-\emph{sn}-glycero-3-phosphoethanolamine, tiethylammonium salt (NBD-PE) was purchased from Molecular Probes (Invitrogen, Copenhagen, Denmark). The sucrose and glucose as well as the calcein were purchased from Sigma (Saint Louis, MS). All the aqueous solutions were prepared in milliQ water.

Giant unilamellar vesicles were prepared via electroformation as described in the literature~\cite{menger1998giant}. First, platinum wire electrodes inserted into Teflon wells are painted with POPC lipids dissolved in chloroform at 2 mg/mL (2-4 $\mu$L) and labeled with 1 mol\% NBD-PE. The electrodes are then allowed to dry in vacuum overnight. 1 mL of sucrose solution (1 Molar) is then added to the wells and the electrodes are connected to an AC signal generator producing a sinusoidal 2V peak to peak signal at 10 Hz for 1.5 hrs. 100 $\mu$L of solution containing vesicles is taken from the well, and the vesicles are then resuspended in 300 $\mu$L of a glucose solution with the same osmolarity as the sucrose solution (1 Molar). This sucrose-glucose system results in the precipitation of the vesicles due to density differences between the glucose and sucrose solutions.

The sedimentation of vesicles was studied through a homebuilt SPIM (UMR5088 Universite Paul Sabatier, Toulouse III and CNRS)~\cite{greger2007basic}. First the glucose solution is placed into a quartz cuvette, followed by careful injection of the sucrose solution containing the vesicles. Once the sample is prepared, precipitating vesicles are localized with a 20X objective in air and a series of images are acquired through a CCD camera with a 15 seconds interval between each image.


\subsection*{Confocal microscopy and image analysis} 
\label{sub:confocal_microscopy_and_image_analysis}

Vesicles at equilibrium were studied using confocal microscopy using an Olympus FV1000 (Tokio, Japan). Calcein was added to the glucose solution at a 1 $\mu$M concentration in order to improve contrast. The calcein provides the external solution with counterions at an approximately 2 $\mu$M concentration.

In order to measure the deformation of the vesicles, the confocal stacks were analyzed using an in-house algorithm by the following method: in each image plane, a threshold was applied, and the boundaries of the vesicle were detected using the function \texttt{cvFindContours} of the OpenCV library~\cite{bradski2000opencv, nelder1965simplex}. The \texttt{cvFitEllipse2} function from the same library was then used on the detected contours to retrieve the dimensions and positions of the vesicles found in the given plane. A KMean algorithm was applied on the vesicle centers to track each vesicle from plane to plane throughout the stack. After this automated detection, the contour of the vesicle along the \emph{z} axis is defined by its borders in each image plane. The Nelder-Mead simplex algorithm~\cite{nelder1965simplex} is then used to fit the \emph{z} contour of the vesicle with a function given by:
\begin{equation}\label{eq:contourfit}
    r = \left(1+\sin^2{\frac{\theta}{2}}\right)^{\alpha}\left(1+\cos^2{\frac{\theta}{2}}\right)^{\beta}
\end{equation}
The center of the polar coordinates in the \emph{z} axis corresponds to the plane where the vesicle width is maximum. Best fit is obtained by finding the best values of parameters $\alpha$ and $\beta$ and minimizing the least-square error. From this fit, area strain and curvature are determined.

To measure the equilibrium distance we waited for 15 minutes after putting the sample to start taking images of the vesicles. After waiting for this period of time no significant movement of the vesicles was observed. The whole experiment typically lasted for an hour. We performed these experiments for unlabeled POPC vesicles and for NBD-PE labeled POPC vesicles in solutions with no added salt and in solutions with 3 mM NaCl concentration.   



\subsection*{Fluid mechanics model} 
\label{sub:fluid_mechanics_model}

\begin{figure}[!ht]
   \begin{center}
     \includegraphics{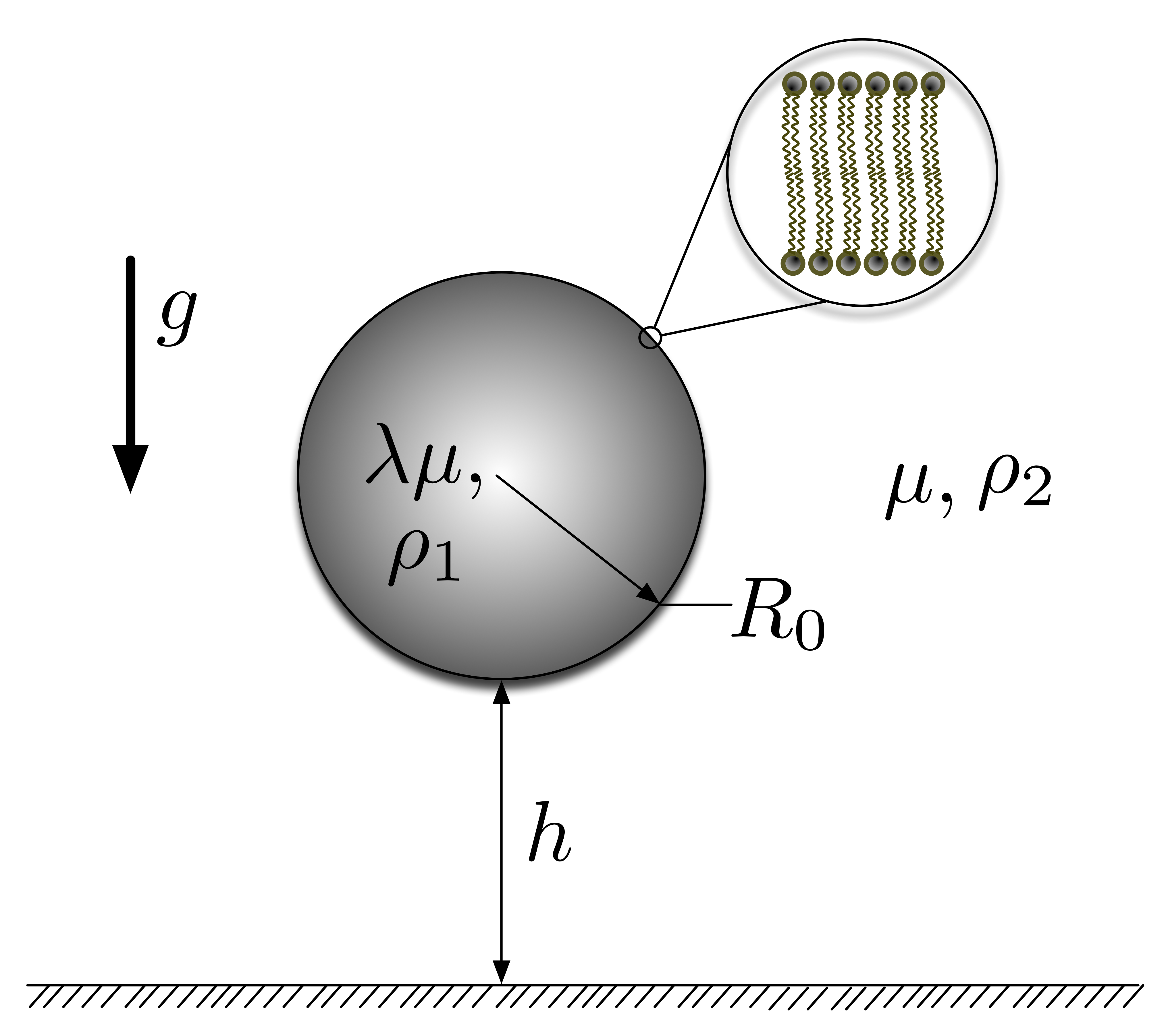}
   \end{center}
   \caption{{\bf Schematic of vesicle subjected to gravity induced sedimentation.} Sedimentation is driven by the density difference $\Delta\rho=\rho_1-\rho_2$.}
   \label{fig:sketch}
\end{figure}

The system considered in this paper is a giant lipid vesicle (diameter $\approx 20 \mu m$) immersed into a semi-infinite external fluid. The vesicle is settling due to a gravitational force towards a no-slip, infinite planar surface. A schematic of the system is shown in Figure~\ref{fig:sketch}. Fluid 1, which is internal to the vesicle, and fluid 2, in which the vesicle is suspended are both assumed to be Newtonian. The fluids are separated by a phospholipid bilayer membrane which introduces a jump in the stress field between both fluids. Gravitational acceleration $g$ is assumed to act along the z-axis, normal to the planar surface. $R_0$, is the radius of the quasi-spherical undeformed vesicle. $\rho_i$ ($i=1,2$) is the density of each fluid. The viscosities of fluids 1 and 2 are $\lambda\mu$ and $\mu$, respectively, where $\lambda$ is the viscosity contrast between the internal and external fluids.

All equations are presented in dimensionless form unless otherwise noted. Distances are scaled using $R_0$. The characteristic time and traction are, respectively,
\begin{equation}\label{eq:scaling}
    t_0 = \frac{\mu (1+\lambda)}{\Delta \rho g R_0},\ \textrm{ and }\ f_0=\frac{\kappa_b}{R_0^3}
\end{equation}
where $\Delta \rho = \rho_1-\rho_2$ and $\kappa_b$ is the bending rigidity modulus of the membrane.

The analysis presented in this work is based upon the creeping motion approximation in which the inertial terms in the equation of motion are neglected entirely for vanishing small Reynolds numbers. In this limit the Stokes equations can be used to model the flow of fluids 1 and 2. During sedimentation the membrane of the vesicle is a moving interface, whose position is unknown \emph{a priori}. The fluid--structure interaction problem, the Stokes equations coupled to the membrane mechanics equations, can be solved using the boundary integral formalism~\cite{pozrikidis:1992tc, zinchenko:1997lk} which yields in non-dimensional form
\begin{equation}\label{eq:boundary-integral}
	u_j = \frac{2}{g_0}\int_s \Delta f_i G_{ij} dS + 2\tilde{\kappa}\int_s u_i T_{ijk} n_k dS
\end{equation}
where $\mathbf{u}$ is the velocity at each node over the membrane, $\tilde{\kappa}=(\lambda-1)/(\lambda+1)$ and $g_0= \Delta \rho g R_0^4/\kappa_b$~\cite{henriksen:2004jq} is analogous to the Bond number. Experiments performed on pure POPC membranes have determined that $\kappa_b$ ranges from 10$k_BT$ to 40$k_BT$~\cite{olsen2009perturbations, henriksen:2004jq} where $k_B$ is the Boltzmann constant and $T$ is absolute temperature. We have chosen the value of 40$k_BT$ as the characteristic value to scale the experimental data and as a parameter in the computer simulations. The traction jump $\mathbf{\Delta f} \equiv {(\mathbf{\sigma^{(1)}}-\mathbf{\sigma^{(2)}})}\cdot \hat{\mathbf{n}}$, where $\hat{\mathbf{n}}$ is the unit vector normal to the surface, is determined from the configuration of the membrane. This term will be further discussed in the following section.

The Green's function kernels $\mathbf{G}$ and $\mathbf{T}$ for a wall bounded semi-infinite space were derived by Blake~\cite{blake1971note} and can be viewed as the fundamental solution to Stokes equations plus additional terms to account for the presence of the planar wall.


\subsection*{Membrane mechanics} 
\label{sub:membrane_mechanics}

The traction jump $\mathbf{\Delta f}$ introduced in Eq.~\ref{eq:boundary-integral} must include all external forces acting on the vesicle. For the problem under consideration $\mathbf{\Delta f} = \mathbf{\Delta f}_{mem} + \mathbf{\Delta f}_{grav} + \mathbf{\Delta f}_{elec}$, where the terms on the right refer to the contribution of the membrane elasticity, the gravitational pull and the electrostatic interactions, respectively. Vesicle deformation at equilibrium results from the balance of all these terms.

There has been extensive theoretical research on lipid membrane deformation that involves the minimization of the free energy functional first proposed by Helfrich~\cite{helfrich:1973eg}. In general, it has been observed that the shape of a vesicle in suspension is limited by its resistance to bending, which is governed by the bending rigidity modulus of the membrane. The shape of the vesicle is determined by minimization of the shape energy which may be written as
\begin{equation}\label{eq:hamilton}
	\mathcal{H}_H = \frac{\kappa_b}{2}\oint(C_1+C_2)^2dS +\sigma\oint dS,
\end{equation}
where $C_1$ and $C_2$ are the principal curvatures of the membrane. Both integrals are over the surface area of the membrane. The last term in Eq.~\ref{eq:hamilton} takes into account the constraint of constant area~\cite{zhong-can:1989zn}. 

It is customary to assume membrane incompressibility based on the high energy cost of surface area dilation of the lipid membrane as compared to other modes of deformation. Most authors rely on approximations based on the calculation of membrane tensions that are proportional to membrane strain~\cite{kaoui2011two, biben:2011vy} which limit deformability of the membrane. Incompressibility can be imposed exactly by explicitly calculating the tension corresponding to the Lagrange multiplier by enforcing a divergence free velocity field on the membrane surface. This has been done for vesicles undergoing small deformations~\cite{seifert:1999sm,boedec2012settling,zhao:2011the-dynamics} and for vesicles in a wall-bound shear flow~\cite{zhao2011dynamics}. An algorithm for the simulation of three dimensional vesicle dynamics was developed by Boedec and coworkers~\cite{boedec20113d}.

Micropipette aspiration experiments have shown that vesicle strain can occur at low tensions~\cite{evans:1990gl}. It has been hypothesized that this strain in the low-tension regime results from the smoothing of thermal membrane fluctuations occurring in the suboptical range, which act as an area reservoir. Area dilation as a function of tension is thus given by~\cite{evans:1990gl}
\begin{equation}\label{eq:evans-rawicz}
	\alpha = \frac{k_BT}{8\pi \kappa_b}\ln{\left(1+c\Sigma A\right)} + \frac{\Sigma}{K_{ext}}.
\end{equation}
where $c$ is a coefficient that depends on the topology of the vesicle, $\Sigma = \sigma R_0^2/\kappa_b$ is the effective reduced tension~\cite{henriksen:2004jq} and $K_{ext}=K_A R_0^2/\kappa_b$ is the dimensionless membrane's modulus for direct area expansion. The variable $A$ is the dimensionless surface area of the membrane. Given the relatively large values of the membrane dilation modulus $K_A$ compared to $\kappa_b$, $K_{ext}$ is a large number of order $~10^8$.

At low tensions, stretching of the membrane is ruled by the thermal energy (logarithmic term in Eq.~\ref{eq:evans-rawicz}) which is characterized by the bending coefficient, while at high tensions membrane strain involves surface dilation and is dominated by $K_{ext}$ (linear term in Eq.~\ref{eq:evans-rawicz}). While membrane strain due to direct surface dilation is difficult to achieve under normal flow conditions due to the relatively high magnitude of $K_{ext}$, smoothing of thermal undulations are easily accessible at the low tensions that arise during gravity induced sedimentation.

In the present work we use Eq.~\ref{eq:evans-rawicz} to calculate the tension on the membrane for a given area strain. Although we are modeling a dynamic process, the sedimentation is slow and will be regarded as a quasi-equilibrium process for which a uniform membrane tension can be assumed. We have performed simulations which have included the calculation of local tensions to enforce local incompressibility and have found the tangential stresses due to tension gradients to be at least two orders of magnitude smaller than the other forces acting on the membrane at any given time for the conditions being considered.


The contribution of the membrane elasticity to the force density can be obtained through the functional derivative of Eq.~\ref{eq:hamilton} as given by~\cite{zhong-can:1989zn}. In non-dimensional form the Helfrich force can be expressed as
\begin{equation}\label{eq:dfmem}
    \mathbf{\Delta f}_{mem} = -\left(\left(4H^3+2\nabla^2_sH-4K_GH\right) + 2\Sigma H\right)\mathbf{\hat{n}}.
\end{equation}
Here, $H=(C_1+C_2)/2$ is the mean curvature, $K_G=C_1C_2$ is the Gaussian curvature and $\nabla^2_s$ is the Laplace-Beltrami operator over the surface of the membrane. Since the contribution of the curvature term to the total energy will depend on the bending modulus, membrane deformation will depend on factors that influence this parameter such as lipid composition~\cite{olsen2009perturbations}.

The gravitational pull is incorporated into the calculations through~\cite{pozrikidis:1992rz} 
\begin{equation}
   \mathbf{\Delta f}_{grav}=-g_0 \mathbf{z},
\end{equation}
where $\mathbf{z}$ is the vertical position of each differential element of the membrane with respect to a reference plane, in this case the glass surface, which has been made dimensionless by scaling with respect to $R_0$.

The calculation of the electrostatic contribution $\mathbf{\Delta f}_{elec}$ is explained in the following section.


\subsection*{Electrostatic interactions} 
\label{sub:electrostatic_interactions}

Experiments performed in the absence of salt result in long screening lengths. For this reason the electrostatic interaction of the vesicle with the surface will play an important role in determining the sedimentation dynamics of the vesicle and the equilibrium state. Microscope glass slides were used as substrates in the experiments. The glass, unless treated using methods such as silanization~\cite{cohen2003electrophoretic}, will present a surface charge, where the main mechanism by which the glass surface acquires charge while in contact with water is by dissociation of the silanol groups~\cite{iler1979chemistry}. The glass surface charge density is negative and has a value around $\sigma_{glass} = -0.2$ mC/m$^2$ for glass in water at pH 7.5~\cite{polin2007colloidal,behrens2001charge}. Vesicles are labeled with 1 mol\% NBD-PE fluorescent probe, which has a negative charge and gives the membrane a negative surface charge density~\cite{groves1998electric}. Hence, the electrostatic interaction between the vesicle and the glass will be repulsive. We can obtain an expression for the interaction force between the membrane and the glass surface from the linearized Poisson-Boltzmann (P-B) equation~\cite{derjaguin:1940mk,carnie1994electrical}
\begin{equation}\label{eq:poisson-boltzamnn}
	\nabla^2 \Psi = \kappa^2 \Psi.
\end{equation}
$\Psi$ is the electric potential and $\kappa^{-1}$ is the Debye length:
\begin{equation}\label{eq:debye}
	\kappa = \left(\frac{2e^2C}{\varepsilon_wk_BT}\right)^{\frac{1}{2}},
\end{equation}
where $e$ is the fundamental charge, $C$ is the ion concentration and $\varepsilon_w$ is the dielectric constant of water. Equation~\ref{eq:poisson-boltzamnn} is valid for potentials less than 40 mV (about twice the thermal potential, $k_bT/e$)~\cite{carnie1994electrical}. In our first set of experiments no salts are added to the solutions, resulting in a Debye length of about 300 nm due to the presence of hydroniums and residual salts. When calcein is added to the solution the Debye length is lowered to 214 nm.The interaction between the vesicle and the glass surface can be modeled using the linear Derjaguin approximation which expresses that the interaction force between a small plane segment of the membrane and the plane surface of the glass is given by~\cite{derjaguin:1940mk}
\begin{equation}\label{eq:derjaguin}
	f(h) = \frac{\varepsilon_w}{8\pi}\frac{\kappa^2\Psi_{ves}\Psi_{glass}}{\cosh{(\kappa h/2)}}
\end{equation}
where $h$ is the distance between the two surfaces and the potentials, for the glass and the vesicle are given, respectively, by:
\begin{equation}\label{eq:potential_glass}
	\Psi_{glass} = \frac{\sigma_{glass}}{\varepsilon_w \kappa}
\end{equation}
and
\begin{equation}\label{eq:potential_ves}
    \Psi_{ves} = \frac{\sigma_{ves} R_0}{(1+\kappa R_0)\varepsilon_w}.
\end{equation}
The surface charge density of the vesicle can be estimated to be around
$\sigma_{ves}\sim -2.9$~mC/m$^{2}$ and its size ranges from 1 $\mu$m to
50 $\mu$m in radius. Thus, the potential generated by the vesicle is high enough that the linearized P-B Eq.~\ref{eq:poisson-boltzamnn} cannot accurately model its behavior. However, for distances greater than the Debye length the behavior of the solutions of the linear and non-linear forms are very similar, provided the surface potential, $\Psi_{ves}$, is replaced by a renormalized surface potential given by~\cite{trizac2002simple,tellez:2011aa}
\begin{equation}\label{eq:potential_renorm}
	\Psi^*_{ves} = \frac{4k_BT}{e}.
\end{equation}
which is independent of $\sigma_{ves}$ and has a value of around $100$ mV in our experimental setup. Substituting Eq.~\ref{eq:potential_renorm} into \ref{eq:derjaguin} for the vesicle potential,  the expression to calculate the electrostatic repulsion becomes
\begin{equation}\label{eq:derjaguin_mod}
	f(h) = \frac{k_B T}{2\pi e}\frac{\Psi_{glass}\kappa^2\varepsilon_w}{\cosh{(\kappa h/2)}}
\end{equation}

Combining Eqs. \ref{eq:potential_glass} and \ref{eq:derjaguin_mod} and expressing in terms of non-dimensional parameters, the contribution of the electrostatic interaction yields:
\begin{equation}\label{eq:deltafelec}
    \mathbf{\Delta f}_{elec} = \frac{\psi_{ie}}{\cosh{(\kappa_{ie}\tilde{h}/2)}} \hat{\mathbf{z}}
\end{equation}
Where we have introduced the dimensionless parameters $\kappa_{ie}=R_0\kappa$ and
\begin{equation}\label{eq:}
    \psi_{ie}=\frac{R_0^2 \kappa_{ie} \sigma_{glass} k_B T}{2\pi e \kappa_b}
\end{equation}
Note that in Eq.~\ref{eq:deltafelec} the distance $\tilde{h}=h/R_0$ is expressed in dimensionless form dividing the distance by the characteristic length $R_0$.

A total of six dimensionless parameters have been introduced, two from the fluid dynamics model, $\tilde{\kappa}$ and $g_0$, two from the electrostatic model, $\kappa_{ie}$ and $\psi_{ie}$, and two from the membrane mechanics model, $K_{ext}$ and the reduced effective tension $\Sigma$. The latter depends on the instantaneous configuration of the membrane and is recalculated at each time step.

When both the external and internal fluids have the same viscosity, $\tilde{\kappa}=0$ which simplifies Eq.~\ref{eq:boundary-integral} by canceling the second integral in the right hand side. Given that we found a very small difference between the viscosities of the sucrose and glucose solutions, we used this simplifying assumption in the solution of the model.

The remaining four dimensionless parameters are calculated in order to match the experimental conditions. The radius $R_0$ of the vesicle together with all other physical parameters determine the values of $\kappa_{ie}$, $\psi_{ie}$, $K_{ext}$ and $g_0$. A range of vesicle radii was chosen to match experimental condition.



\subsection*{Numerical method} 
\label{sec:numerical_method}

The general boundary integral method has been widely used to model the dynamic behavior of drops, capsules, vesicles and cells suspended in general flows. We refer to~\cite{pozrikidis:1992tc,pozrikidis:2001hs,zinchenko:1997lk} for the general description of the method. However, in the following paragraphs we discuss particular considerations for the simulation of gravity-induced sedimentation.

\begin{figure}[!ht]
   \begin{center}
     \includegraphics{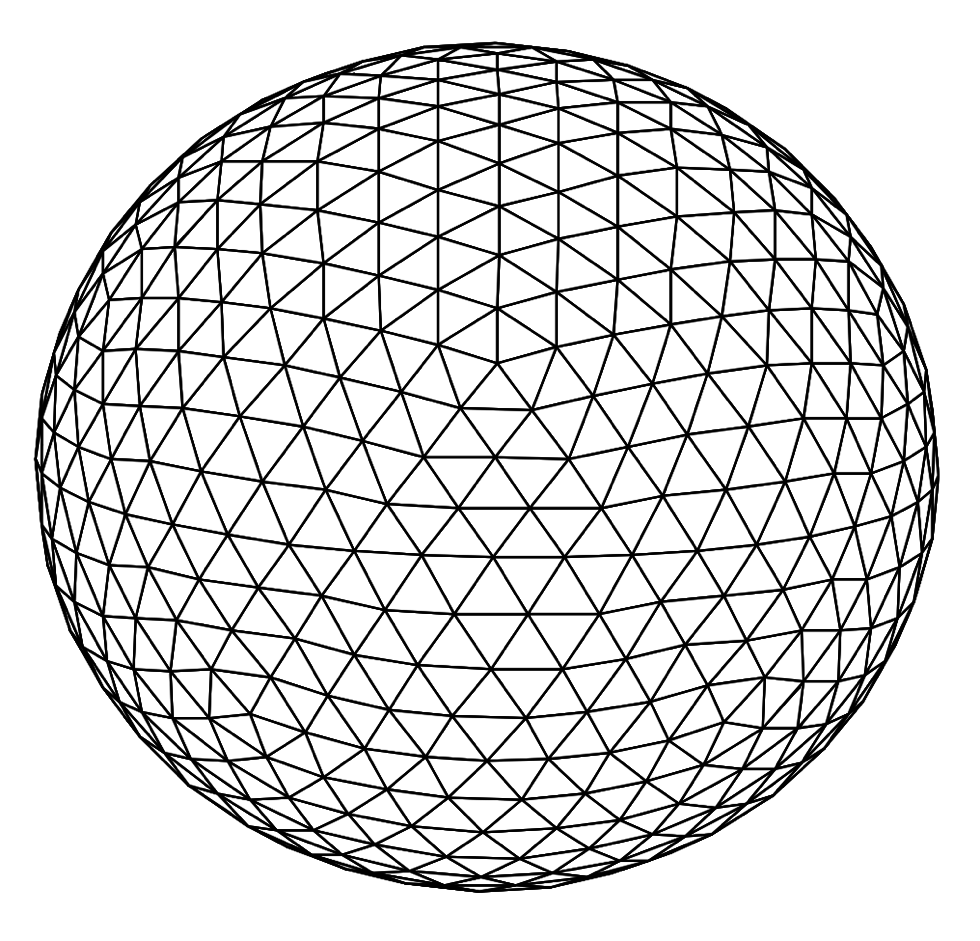} 
   \end{center}
   \caption{{\bf Example of the mesh used in the initial configuration.} The mesh is constructed by refining the triangular faces of an icosahedron inscribed into a sphere by dividing recursively each triangle into four smaller triangles and projecting the resulting nodes to the surface of the sphere. The mesh used in the current work is the result of four refinements in order to obtain a total of 642 nodes and 1280 triangular elements.}
   \label{fig:mesh}
\end{figure}

To obtain the computational domain we use the method of uniform triangulation to discretize the initially spherical membrane into a uniform triangulated mesh. As it has been described by other authors~\cite{zinchenko:1997lk} we begin with a regular icosahedron inscribed into the sphere. Each face of the icosahedron is divided into four smaller triangles by dividing each edge at its midpoint, the new vertices are projected radially onto the sphere and the process is applied recursively as many times as necessary to obtain the desired refinement. In the current work we do three refinements to obtain 642 vertices and 1280 triangular elements. An example of the computational mesh is shown in Fig.~\ref{fig:mesh}.

Integration of the right hand side terms in Eq.~\ref{eq:boundary-integral} are approximated by a simple surface trapezoidal rule that requires the integrands only at the triangle vertices~\cite{zinchenko:1997lk}. This rule can be written, for any function $g(\mathbf{x})$, in the following form, by reassigning the contribution of each triangular element to the vertex:
\begin{equation}
   \int g(\mathbf{x}) \approx \sum_i g(\mathbf{x}) \Delta S_i,
\end{equation}
\begin{equation}\label{eq:sum_int}
   \Delta S_i = \frac{1}{3}\sum S,
\end{equation}
where the summation in (\ref{eq:sum_int}) is over all flat triangular areas $\Delta S$ with vertex $\mathbf{x}_i$.

One of the greatest challenges in the numerical algorithm is the calculation of the curvatures and the Laplace-Beltrami operator in Eq.~\ref{eq:dfmem} over the discrete triangulated surface. A common algorithm used to calculate the mean and Gaussian curvature is to fit via least-squares a quadratic surface to each node and its neighbors and calculate the curvature from this function. From our tests we have determined that this method introduces sufficient error in the calculation of the curvature to be problematic during the calculation of the Laplace-Beltrami operator. For this reason we have decided to use the discrete differential-geometry operators as presented in~\cite{meyer:2002vv}.

\begin{figure}[!ht]
    \begin{center}
      \includegraphics{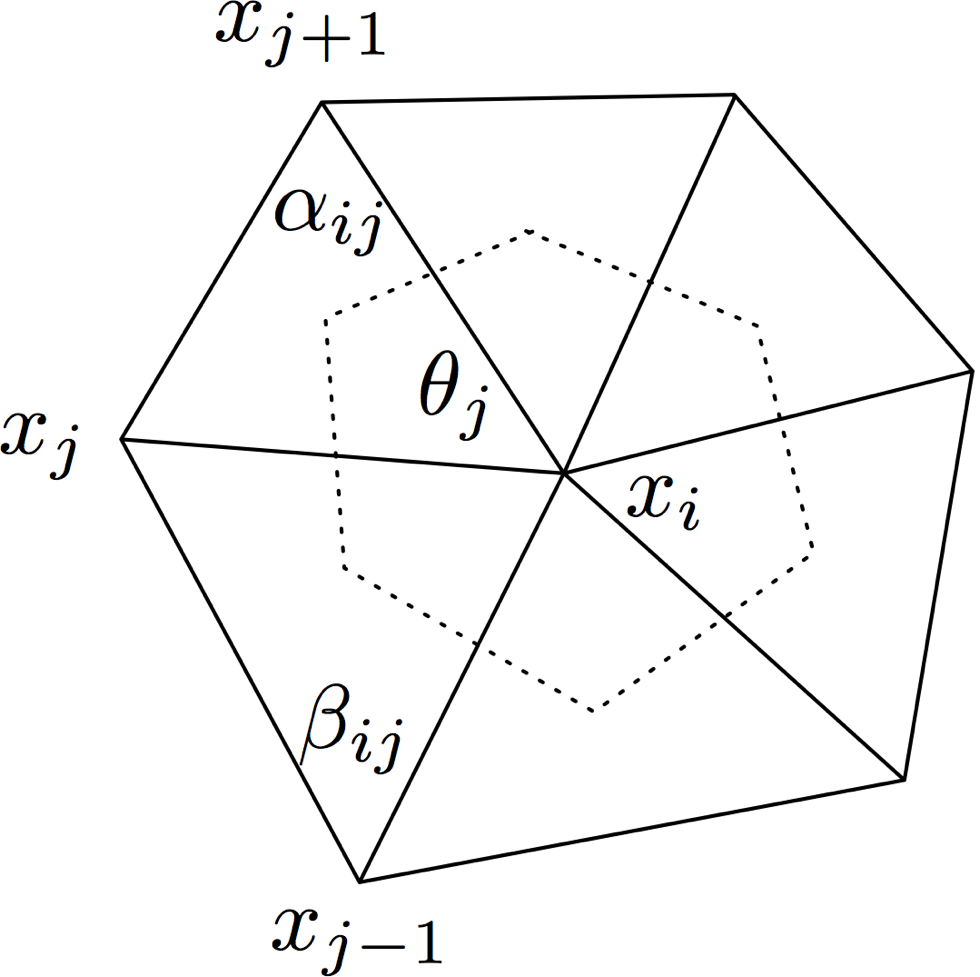}
    \end{center}
    \caption{{\bf 1-ring neighborhood of vertex indicating the sub-area used for computation using the method of Meyer \emph{et. al.}}}
    \label{fig:angles_discrete_ops}
\end{figure}

The mean curvature estimate is derived from a discretization of the Laplace-Beltrami operator applied to the 1-ring neighborhood. Given a patch of triangles surrounding point $\mathbf{x}_i$ as shown in Fig.~\ref{fig:angles_discrete_ops}, the estimates for the Gaussian curvature, $K_i$ and mean curvature $H_i$, at $\mathbf{x}_i$, given by Meyer et. al.~\cite{meyer:2002vv} are:

\begin{equation}\label{eq:gaussian_curv}
    K_i = \frac{1}{A_{\mathrm{Voronoi}}}\left(2\pi-\sum_j\theta_j\right)
\end{equation}
\begin{equation}\label{eq:mean_curv}
    2H_i\mathbf{\hat{n}_i} = \frac{1}{A_{\mathrm{Voronoi}}}\sum_i\left(\cot{\alpha_{ij}}+\cot{\beta_{ij}}\right)\left(\mathbf{x}_i-\mathbf{x}_j\right)
\end{equation}
where $\mathbf{x}_j$, $\theta_j$, $\alpha_{ij}$ and $\beta_{ij}$ are shown in the figure. The area $A_{\mathrm{Voronoi}}$ corresponds to the ``Voronoi area'', defined in each triangle by the point $\mathbf{x}_i$, the midpoints of the triangle edges, and the circumcenter of the triangle, summed over all the triangles. Meyer \emph{et. al.} demonstrated that the error in the computation of the curvature is minimized by using this area, as opposed to other alternatives. The Voronoi area can be calculated by:
\begin{equation}\label{eq:voronoi}
    A_{\mathrm{Voronoi}} = \frac{1}{8}\sum_j\left(\cot{\alpha_{ij}}+\cot{\beta_{ij}}\right)\|\mathbf{x}_i-\mathbf{x}_j\|^2
\end{equation}

Finally the Laplace-Beltrami operator of the mean curvature $H$ can be estimated as~\cite{meyer:2002vv}
\begin{equation}\label{eq:lablace-beltrami}
    \nabla^2_sH = \frac{1}{A_{\mathrm{Voronoi}}}\sum_{j \in N_1(i)} \left(\cot{\alpha_{ij}} + \cot{\beta_{ij}}\right)\left(H_j-H_i\right)
\end{equation}
A similar method was recently used by Boedec and coworkers~\cite{boedec20113d} for the calculation of the curvatures and Laplace-Beltrami operator.

The numerical algorithm advances in a relatively simple fashion and can be described as follows: from the shape of the vesicle the traction $\Delta f$ on the membrane is computed, the boundary-integral Eq.~\ref{eq:boundary-integral} can then be solved for the velocity at the nodes, finally nodes are advected via integration in time using a second order Runge-Kutta method. The algorithm continues until an equilibrium state is achieved. 

An important difficulty in the boundary-integral calculations is efficient mesh control. Namely, if the nodes are simply advected with the flow, an initially regular mesh of triangles covering the surface becomes highly irregular after a short simulation time, thus invalidating the calculation. Mesh degeneration is especially problematic in gravity-induced motion given the large displacements of the vesicles. A mesh stabilization method can be developed by, instead of advecting the membrane nodes with the interfacial velocity $\mathbf{u}$ from Eq.~\ref{eq:boundary-integral}, using the normal velocity $(\mathbf{u}\cdot\mathbf{\hat{n}})\mathbf{\hat{n}}$ plus an artificial tangential velocity field which can be constructed to maintain certain uniformity of the mesh. In the current work we construct the tangential velocity field using the passive mesh stabilization algorithm first introduced by Zinchenko and coworkers \cite{zinchenko:1997lk}. The algorithm is based on the \emph{global} minimization of the rate of change of the distances between neighboring nodes. To achieve this one must solve an optimization problem over the whole surface of the membrane. We found that by using such algorithm long simulations of gravity-induced sedimentation of vesicles were possible with reasonable time steps.



\section*{Results and Discussion} 
\label{sec:results_and_discussion}

In this section the results from the computer simulations are presented and compared to the experimental data. First, the sedimentation dynamics are studied, then the equilibrium state of the vesicle is analyzed.

\subsection*{Sedimentation of vesicles towards a flat surface} 
\label{sub:sedimentation_of_vesicles_towards_a_flat_surface}

\begin{figure}[!ht]
    \begin{center}
       \includegraphics{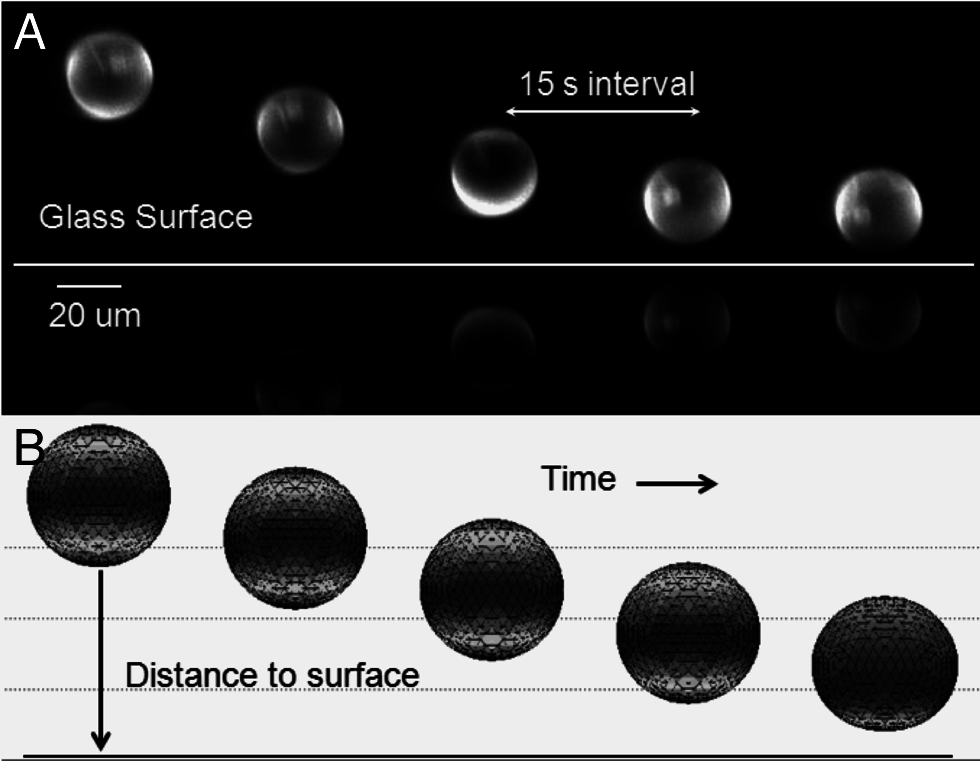}
    \end{center}
    \caption{{\bf Image sequence for the sedimentation of a vesicle.} The time interval between images is 15 s. A. Images obtained through SPIM for $g_0=3$. B. Images from computer simulation depicting sedimentation of a vesicle under similar conditions.}
    \label{fig:montage}
\end{figure}

The sedimentation of vesicles in isotonic conditions was studied using SPIM. The sedimentation process is driven by the density difference between the sucrose and glucose solutions ($\approx$35 kg/m$^3$). The best six image series were analyzed. The criterion for this selection was to ensure that the vesicle did not present significant lateral displacement during sedimentation. Fig.~\ref{fig:montage} shows sample images from one of the series that were analyzed. The sedimentation rate from the images is calculated by a simple two point backwards finite difference scheme for the position of the centroid between successive images. All experimental results have a time lapse between images of 15 seconds.

Due to the optical configuration of the SPIM, an interference pattern results in an image artifact where some regions of the membrane appear to be brighter. This could be interpreted as these regions having a higher concentration of fluorescente probe. Nevertheless, given the preparation technique and the experimental configuration we do not expect this to be the case. We consider the assumption of uniformly distributed probe to be more accurate. 

The figure also shows the sedimentation sequence as predicted by our computer simulations under similar conditions. Computationally, we calculate the sedimentation rate with a method similar to the one used experimentally, but the time between data points is much smaller. The simulations are initialized with the vesicle at a certain distance from the surface (between 6-8 radii). The position of the centroid of the vesicle is calculated at each time step and the sedimentation rate is calculated.

\begin{figure}[!ht]
    \begin{center}
      \includegraphics{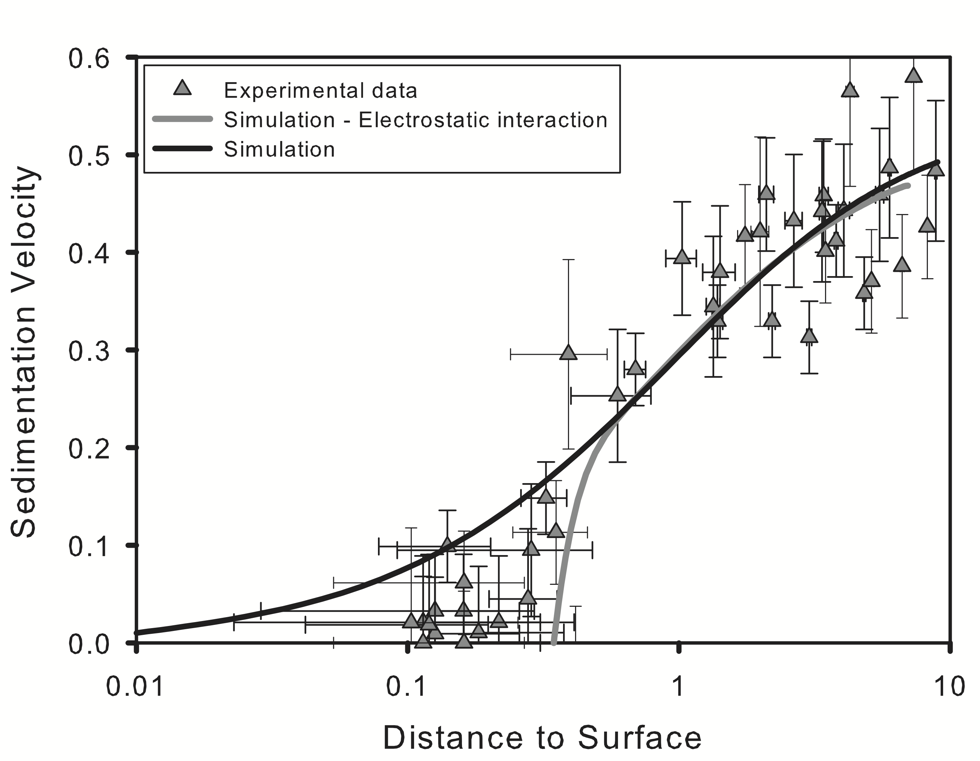}
    \end{center}
    \caption{{\bf Sedimentation rate as a function of the distance to the surface.} Experimental data (symbols) and computer simulations (lines) are shown. The light solid line corresponds to a simulation which incorporates the electrostatic interaction between the vesicle and the surface, the dark solid line corresponds to a simulation which does not take into account this interaction. }
    \label{fig:sedimentation}
\end{figure}

Sedimentation rates calculated experimentally and computationally are shown in Fig.~\ref{fig:sedimentation}. In the figure the distance to the surface is set to non-dimensional units by dividing by $R_0$ and velocity is also given in non-dimensional form. Far away from the surface, the simulations match the experimental results accurately. The sedimentation velocity decreases as the vesicle approaches the flat surface and the rate of change is well predicted by the simulations. At distances larger than $0.2$ the electrostatic interactions are negligible and the whole process is dominated by the hydrodynamics.

As the vesicle approaches the wall, it decelerates. For uncharged vesicles (dark solid line in the figure) the velocity decreases exponentially approaching zero. This behavior is a result of the slow draining of the fluid that separates the vesicle and the surface, which corresponds to a lubrication regime~\cite{ascoli:1990xy}. When the electrostatic interaction is taken into account (light solid line in the figure) the sedimentation velocity decreases rapidly to zero further away from the surface avoiding this regime entirely. Close to the wall the electrostatic interactions
become increasingly important causing a rapid deceleration, as the
gravity-induced force is balanced by the electrostatic
interactions. 

Closer to the surface there appears to be a larger discrepancy between the experimental and computational results. Experimental data lies between the simulations with and without electrostatic interactions. We attribute this to the method used in determining the distance between the vesicle and the surface in the SPIM experiments, in which the gap between the vesicle and the surface is determined by calculating the distance between the centroids of the vesicle and its reflection (see Fig.~\ref{fig:montage}) divided by two, minus the radius of the vesicle. This method does not take into account the deformation of the lower part of the vesicle, which induces an underestimation of the size of the gap at equilibrium.

One could also look for an explanation to this discrepancy in the assumptions made when developing our model. The most important simplification affecting the sedimentation rate is the assumption of having a uniform tension on the membrane. Given the symmetry of the problem, during the sedimentation process the membrane will become immobile and flow inside of the vesicle will be restricted. With a uniform tension, the membrane is mobile and there is the possibility of flow inside of the vesicle, as it happens in sedimenting droplets. This phenomenon will directly affect sedimentation rate. We have performed simulations, and compared the velocity of particles with mobile and immobilized interfaces under the same conditions and have found differences of less than 0.1 percent under these conditions. This difference is well below our experimental uncertainty. Also, this effect would be more important far away from the wall where the sedimentation rate is higher, opposite to our observations. Finally, in the following section we show that our model accurately predicts the equilibrium position of the vesicle as measured by the more precise confocal microscopy. 

Once the forces acting on the vesicle balance out, the vesicle reaches equilibrium and the sedimentation rate approaches zero. At this point the vesicle acquires a shape determined by the equilibrium of the electrostatic and gravitational forces with the membrane resistance to deform.


\subsection*{Gap between the vesicle and the glass surface at equilibrium} 
\label{sub:gap_between_the_vesicle_and_the_glass_surface_at_equilibrium}

The gap between the vesicle and the surface at equilibrium was further studied using confocal microscopy images. Confocal images and the corresponding frame from the computer simulation for similar conditions are shown in Fig.~\ref{fig:equilibrium}. The small gap between the vesicle and the surface is seen for both the confocal image and the computer simulation.

The gap at equilibrium measured experimentally and in the computer simulations are shown in Fig.~\ref{fig:equilibrium_distance} as a function of $g_0$. The Debye length for NBD-PE labeled vesicles was calculated to be 214 nm when the ionic concentration of the calcein is accounted for. Computed simulations using that value for the Debye length show good agreement with the experimental results. The confocal microscopy combined with our image analysis algorithm provides a more accurate way of measuring the gap than the SPIM images. The distance to the surface decreases as $g_0$ increases due to the increase in the effective weight of the membrane as compared to the electrostatic force. 

Experimentally, the addition of a small amount of salt results in the screening of the electrostatic forces and a decrease of the Debye length. By adding 3 mM of NaCl the Debye length is decreased to 5.5 nm. This condition can be used as a control, to verify that our observations are due to the electrostatic interactions. The experimental data and computer simulations under these conditions are also shown in Fig.~\ref{fig:equilibrium_distance}. Overall, under these conditions, we observe a decrease of the equilibrium distance between the vesicle and the surface. The computer simulations predict a smaller gap distance than those measured experimentally, especially for smaller vesicles. Before commenting on the possible explanations for this behavior we note in passing that for larger vesicles the gap distance seems to converge for both Debye lengths being considered. This indicates that for heavier vesicles, which can reach closer to the wall, sedimentation is limited by the viscous draining of the fluid between the vesicle and the wall rather than the electrostatic interactions.


The gap size is determined through image analysis using a custom ImageJ~\cite{abramoff:2004fk} algorithm. The process consists in plotting the intensity profile along a line that crosses the water gap between the vesicle and the glass surface. From the plot, the gap size is defined as the distance between points with an intensity corresponding to 70 percent of the highest measured value. Several distances measured for the same vesicle (between 3 and 5) are then averaged. The error bars shown in figures 7 and 8 are obtained from a combination of the deviation from the mean of the measured values with a confidence interval of 95\%, together with the bias error due to the resolution of the microscope configuration used (around 0.29 microns).

Our scaling normalizes the weight of all vesicles. It can be observed in Fig.~\ref{fig:sedimentation} that sedimentation rate is independent of the size of the vesicle in non-dimensional terms. Physically, sedimentation rate does depend on the size of the vesicle, as heavier vesicles sediment faster. From the characteristic length $R_0$ and the characteristic time defined in Eq.~\ref{eq:scaling} it can be seen that physical velocity varies with $R_0^2$. That is, vesicles with radius of order one are sedimenting a hundred times slower that those of radius 10! Due to the stability of the vesicles, experiments are limited to one hour. With this in mind, a better explanation to the observed behavior is that smaller vesicles have not yet reached equilibrium and are sedimenting very slowly when the measurement is performed. Computationally, we are allowing the vesicles to reach equilibrium defined as the state when displacements are below the numerical resolution of our algorithm, thus the measured gap is smaller.


Overall, the computational model is able to predict in an accurate fashion the size of the gap at equilibrium for a wide variety of conditions. The agreement in the results also indicates an appropriate selection of the values of some important parameters that were not measured in our experiments but relied on calculated values and theoretical predictions, such as the glass surface charge density and the counterions concentration in the solutions.

So far we have presented results corresponding to charged vesicles labeled with NBD-PE in solution. We have considered solutions with no added salt where electrostatic interactions are important (relatively large Debye length) and with salt added to screen the effect of electrostatic interactions. The addition of a small amount of salt (3 mM) generates a slightly hypertonic condition which causes the vesicle to deflate. In order to investigate the effect of this on the measured gap we perform additional experiments with unlabeled vesicles (no charge). The gap distance for charged vesicles in a solution with added salt compared to unlabeled vesicles is shown in Fig.~\ref{fig:salt_vs_noNBD}. It can be seen that there is no difference in the behavior of both groups and that the same behavior discussed above is observed in unlabeled vesicles. In the following section we investigate the deformation of both charged and uncharged vesicles.

The thorough understanding of the role played by the electrostatic interactions on the sedimentation process, as well as on determining its position at equilibrium might be important in understanding the formation and stability of certain colloidal systems. The results presented in this section will play an important role in this understanding.


\subsection*{Vesicle deformation at equilibrium} 
\label{sub:vesicle_deformation_at_equilibrium}

\begin{figure}[!ht]
    \begin{center}
      \includegraphics{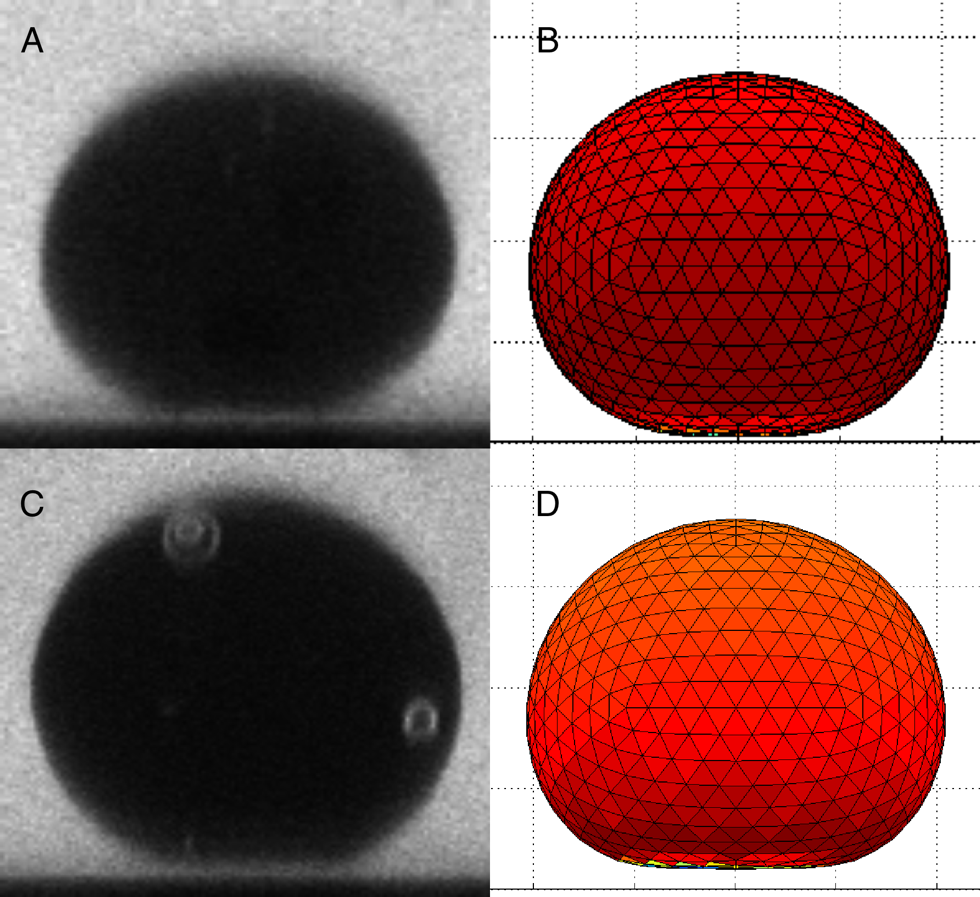}
    \end{center}
    \caption{{\bf Images of vesicles at equilibrium} A. and C. show confocal images of giant unillamellar vesicle at equilibrium for solutions with 3 mM of NaCl and no salt, respectively. Staining the outer solution with calcein (1.0 $\mu$M) allows to visualize the gap between the vesicle and the glass surface. B. and D. show computer simulation of a vesicle at equilibrium under similar conditions as in A. and C., respectively.}
    \label{fig:equilibrium}
\end{figure}

\begin{figure}[!ht]
    \begin{center}
      \includegraphics{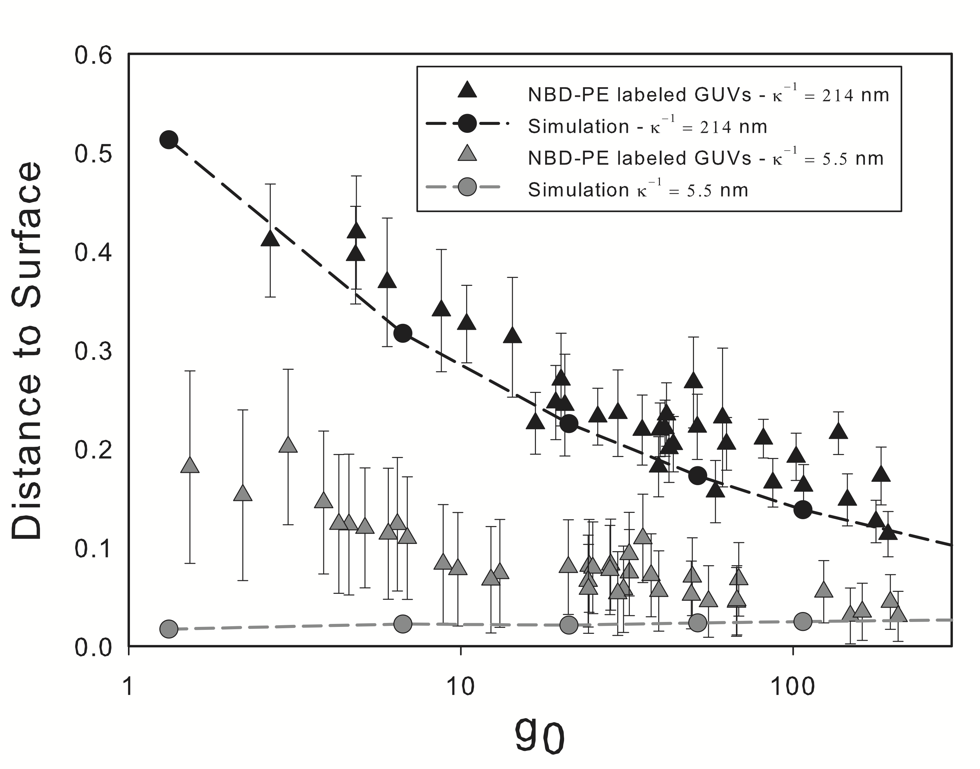}
    \end{center}
    \caption{{\bf Equilibrium distance to the surface.} Gap at equilibrium as a function of $g_0$ obtained from simulations (dashed lines) and experimental data (symbols) for two different Debye lengths ($\kappa^{-1}$). In the simulations $\kappa^{-1}$ is an input parameter, experimentally it depends on the ionic concentration which can be changed by the addition of salt.}
    \label{fig:equilibrium_distance}
\end{figure}

\begin{figure}[!ht]
    \begin{center}
      \includegraphics{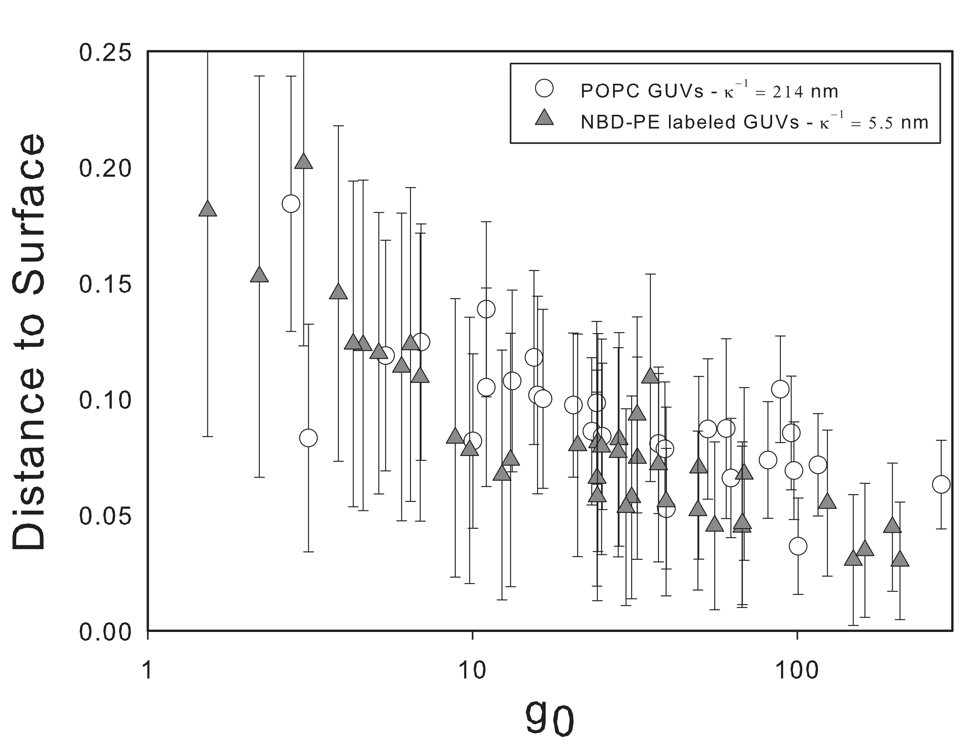}
    \end{center}
    \caption{{\bf Comparison between the effect of significantly screening the electrostatic interaction and absence of charge on the membrane.} Open circles represent experimental data for the equilibrium distance between the vesicle and the surface for non-charged vesicles, while filled triangles show the distance to the surface at which charged vesicles reach equilibrium when 3 mM of salt is added.}
    \label{fig:salt_vs_noNBD}
\end{figure}

When the vesicle reaches equilibrium, the sedimentation process stops and the vesicle adopts an equilibrium shape determined by the balance of the forces acting on it. The shape of the vesicle at equilibrium was studied through confocal microscopy and image analysis as explained in the Materials and Methods section. The analysis was applied to images similar to that shown in Fig.~\ref{fig:equilibrium} (A). Contrast was improved by staining the glucose solution with low concentrations of calcein (1 $\mu$M). At these low concentrations, it should not influence the osmotic conditions of the solution.

\begin{figure}[!ht]
    \begin{center}
      \includegraphics{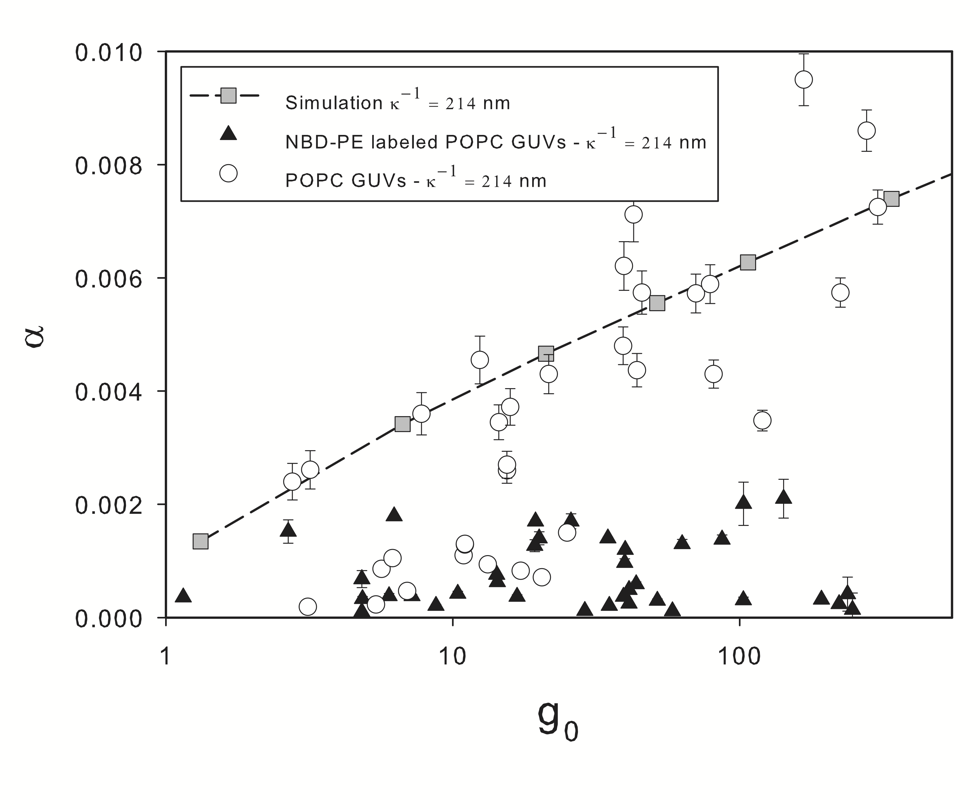}
    \end{center}
    \caption{{\bf Membrane strain at equilibrium measured from experiments and simulations as a function of $g_0$.} Open circles and filled triangles show the measured strain for non-charged and charged vesicles, respectively. Computer simulation results are shows as filled squares and dashed line. The presence of charge in the membrane appears to stiffen the membrane resulting on smaller measured strains.}
    \label{fig:strain_vs_g0}
\end{figure}

It has been shown~\cite{evans:1990gl} that membrane deformation is caused by the superposition of two modes: smoothing of thermal undulations and direct stretching of the membrane. At low external forces, such as those in the present experiments, it is expected that strain be due to the former mode only. Tension on the membrane at equilibrium is proportional to the value of $g_0$. By plotting the area strain as a function of $g_0$ in a semi-log plot (Fig.~\ref{fig:strain_vs_g0}) it is evident that the logarithmic term in Eq.~\ref{eq:evans-rawicz} is the dominating term in defining the shape of the membrane at equilibrium as expected.

In Fig.~\ref{fig:equilibrium} experimental images of deformed vesicles at equilibrium were shown together with the computational prediction at similar conditions. It is interesting to note that those relatively large deformations are achieve with very small area strains (less than 0.01 in all cases), which can be explained due to the smoothing of thermal undulations~\cite{ly:2004fh}. Hence, it becomes apparent that considerations of thermal undulations is important in the study of vesicle sedimentation. 

The tension--strain relationship (Eq.~\ref{eq:evans-rawicz}) suggests that strain at a certain tension depends on the membrane bending rigidity. It has been suggested previously~\cite{rowat:2004fk,mitchell:1989aa,winterhalter:1988wj} that electrostatic interactions between neighboring NBD-PE molecules should provide a contribution to the bending rigidity. In order to verify this observation we compare the strain of charged and uncharged vesicles. 

As mentioned above, salt produces a slightly hypertonic condition which causes the vesicle to deflate, this will certainly affect the deformability of the particles. For this reason in the following we report the deformation of both labeled and unlabeled vesicles suspended in the same solution with no salt added. In the previous section it was shown that not labeling the vesicles had the same effect on the measured gap as adding a small amount of salt to screen the electrostatic interactions. 

Fig.~\ref{fig:strain_vs_g0} shows the measured area strain for both NBD-PE labeled and unlabeled vesicles. The simulation results for vesicles with a bending rigidity modulus $\kappa_b = 40k_BT$ is also shown in the figure. For the majority of the unlabeled vesicles the measured strain is in line with the prediction of the computational model. For a smaller fraction of the data points, mostly lighter vesicles, a much smaller strain is measured. All data points correspond to the same sample, hence tonicity effects should be ruled out. The most probable cause for this observation is, as explained above, that this vesicles have not yet reached an equilibrium state and have not yet fully deform to balance the gravitational force. 

Charged vesicles undergo a smaller strain at equilibrium. Solid triangular data points in Figs.~\ref{fig:equilibrium_distance} and \ref{fig:strain_vs_g0} correspond to the same vesicles. Based on the comparison between the experimental data and the computer simulations, we consider that these vesicles have reached equilibrium, and electrostatic forces are balanced with the gravitational force. These results suggest that the presence of charged fluorescent probes does have the effect of increasing the bending rigidity as has been reported previously~\cite{rowat:2004fk}. 

Measurement of the mechanical properties of vesicles represents one of the possibilities presented by the current computational model. By comparing good experimental data with the computer simulations, one might be able to determine the mechanical properties of membranes which could complement techniques such as micropipette aspiration.



\section*{Conclusions} 
\label{sec:conclusions}

We have reported the implementation of a computational algorithm for the simulation of lipid vesicles in suspension. Experiments using SPIM and confocal microscopy were performed to verify the performance of the algorithm. The algorithm is used to simulate the sedimentation of vesicles due to gravity, taking into consideration the electrostatic interactions between the membrane and a glass surface towards which it is sedimenting. We have shown that our simulations are able to predict, with reasonable accuracy, the rate at which the vesicles sediment and the velocity variations as the vesicle approach the glass surface. The algorithm also predicts the equilibrium gap between the vesicle and the surface at equilibrium. It was shown that the consideration of the electrostatic contribution to surface interactions is essential in order to accurately predict the sedimentation rate, especially at close range from the surface, and the fluid gap between the vesicle and the surface at equilibrium.  

We have shown that by modeling the mechanical behavior of the membrane with Eq.~\ref{eq:evans-rawicz}, which superposes the deformation due to the smoothing of thermal undulations and direct membrane stretch, area strains are reasonably small and in line with those observed experimentally. The model used in our simulations also takes into account the intrinsic mechanical properties of the membrane through the membrane bending modulus, which can be modulated in cells through membrane composition. For example, by adding 30\% cholesterol the bending rigidity of the membrane can be doubled~\cite{pan2008cholesterol}. It is also suggested that charged fluorescent probes on the membrane have the effect of increasing its bending rigidity.

Our vesicle model has considered the basic physical contributions: Stokes flow, bending rigidity, membrane deformation through smoothing of thermal undulations but limited by the high energy cost of direct membrane deformation and electrostatic interactions between the vesicle and the glass surface. In its current form the algorithm can be used to study the effect that the different physical parameters discussed have on the sedimentation rate, distance to the surface at equilibrium and membrane strain. It can also be extended in order to study important phenomena such as colloidal stability, and biofilm formation.

Finally, regarding future work, we consider our current model to be robust and accurate enough to allow us to investigate the roll that the physics described and studied in the current paper play in the slow motion of a suspended vesicle. Nevertheless, the dynamics and motion of vesicles under shear or general flow pose increasing challenges as the calculation of local tensions becomes mandatory. The method proposed by Boedec \emph{et al.} is a very good starting point, but the inclusion of viscosity contrasts different than one and other physics, such as those investigated in the current paper, could prove challenging and to the best of our knowledge remain unsolved.

\section*{Acknowledgments}

We acknowledge Brice Ronsin from the Toulouse RIO Imagin facility (CNRS UMR 5547 Universit\'e Paul Sabatier, Toulouse) and the MEMPHYS group from Southern Denmark University for their help in the development of the experimental procedures of this work.


\begin{thebibliography}{10}
\providecommand{\url}[1]{\texttt{#1}}
\providecommand{\urlprefix}{URL }
\expandafter\ifx\csname urlstyle\endcsname\relax
  \providecommand{\doi}[1]{doi:\discretionary{}{}{}#1}\else
  \providecommand{\doi}{doi:\discretionary{}{}{}\begingroup
  \urlstyle{rm}\Url}\fi
\providecommand{\bibAnnoteFile}[1]{%
  \IfFileExists{#1}{\begin{quotation}\noindent\textsc{Key:} #1\\
  \textsc{Annotation:}\ \input{#1}\end{quotation}}{}}
\providecommand{\bibAnnote}[2]{%
  \begin{quotation}\noindent\textsc{Key:} #1\\
  \textsc{Annotation:}\ #2\end{quotation}}
\providecommand{\eprint}[2][]{\url{#2}}

\bibitem{tomaiuolo:2011us}
Tomaiuolo G, Guido S (2011) Start-up shape dynamics of red blood cells in
  microcapillary flow.
\newblock Microvascular research 82: 35--41.
\input{tomaiuolo:2011us}

\bibitem{de2007plasma}
De~Rosa M, Alinovi C, Galtieri A, Scatena R, Giardina B (2007) The plasma
  membrane of erythrocytes plays a fundamental role in the transport of oxygen,
  carbon dioxide and nitric oxide and in the maintenance of the reduced state
  of the heme iron.
\newblock Gene 398: 162--171.
\input{de2007plasma}

\bibitem{yoon2009flickering}
Yoon Y, Hong H, Brown A, Kim D, Kang D, et~al. (2009) Flickering analysis of
  erythrocyte mechanical properties: dependence on oxygenation level, cell
  shape, and hydration level.
\newblock Biophysical journal 97: 1606--1615.
\input{yoon2009flickering}

\bibitem{vlahovska:2009fe}
Vlahovska PM, Podgorski T, Misbah C (2009) Vesicles and red blood cells in
  flow: From individual dynamics to rheology.
\newblock Comptes Rendus Physique 10: 775--789.
\input{vlahovska:2009fe}

\bibitem{kaoui:2009sx}
Kaoui B, Farutin A, Misbah C (2009) Vesicles under simple shear flow:
  Elucidating the role of relevant control parameters.
\newblock Physical Review E 80: 61905.
\input{kaoui:2009sx}

\bibitem{kaoui:2009rm}
Kaoui B, Biros G, Misbah C (2009) Why do red blood cells have asymmetric shapes
  even in a symmetric flow?
\newblock Physical review letters 103: 188101.
\input{kaoui:2009rm}

\bibitem{helfrich:1973eg}
Helfrich W (1973) Elastic properties of lipid bilayers: theory and possible
  experiments.
\newblock Z Naturforsch 28: 693--703.
\input{helfrich:1973eg}

\bibitem{seifert:1995ui}
Seifert U (1995) The concept of effective tension for fluctuating vesicles.
\newblock Zeitschrift F{\"u}r Physik B 97: 299--309.
\input{seifert:1995ui}

\bibitem{seifert:1997zc}
Seifert U (1997) Configurations of fluid membranes and vesicles.
\newblock Advances in Physics 46: 13--137.
\input{seifert:1997zc}

\bibitem{biben:2011vy}
Biben T, Farutin A, Misbah C (2011) Three-dimensional vesicles under shear
  flow: Numerical study of dynamics and phase diagram.
\newblock Physical Review E 83: 31921.
\input{biben:2011vy}

\bibitem{sukumaran:2001py}
Sukumaran S, Seifert U (2001) Influence of shear flow on vesicles near a wall:
  A numerical study.
\newblock Physical Review E 6401.
\input{sukumaran:2001py}

\bibitem{misbah:2006jl}
Misbah C (2006) Vacillating breathing and tumbling of vesicles under shear
  flow.
\newblock Physical Review Letters 96: 28104.
\input{misbah:2006jl}

\bibitem{kaoui:2008ph}
Kaoui B, Ristow GH, Cantat I, Misbah C, Zimmermann W (2008) Lateral migration
  of a two-dimensional vesicle in unbounded poiseuille flow.
\newblock Physical Review E 77: 21903.
\input{kaoui:2008ph}

\bibitem{kaoui2011two}
Kaoui B, Harting J, Misbah C (2011) Two-dimensional vesicle dynamics under
  shear flow: Effect of confinement.
\newblock Physical Review E 83: 066319.
\input{kaoui2011two}

\bibitem{boedec20113d}
Boedec G, Leonetti M, Jaeger M (2011) 3d vesicle dynamics simulations with a
  linearly triangulated surface.
\newblock Journal of Computational Physics 230: 1020--1034.
\input{boedec20113d}

\bibitem{zhao2011dynamics}
Zhao H, Spann A, Shaqfeh E (2011) The dynamics of a vesicle in a wall-bound
  shear flow.
\newblock Physics of Fluids 23: 121901--121901.
\input{zhao2011dynamics}

\bibitem{abkarian:2005ib}
Abkarian M, Viallat A (2005) Dynamics of vesicles in a wall-bounded shear flow.
\newblock Biophysical Journal 89: 1055--1066.
\input{abkarian:2005ib}

\bibitem{kantsler:2005zo}
Kantsler V, Steinberg V (2005) Orientation and dynamics of a vesicle in
  tank-treading motion in shear flow.
\newblock Physical Review Letters 95: 258101.
\input{kantsler:2005zo}

\bibitem{kantsler:2006lq}
Kantsler V, Steinberg V (2006) Transition to tumbling and two regimes of
  tumbling motion of a vesicle in shear flow.
\newblock Physical Review Letters 96: 36001.
\input{kantsler:2006lq}

\bibitem{deschamps:2009tt}
Deschamps J, Kantsler V, Steinberg V (2009) Phase diagram of single vesicle
  dynamical states in shear flow.
\newblock Physical review letters 102: 118105.
\input{deschamps:2009tt}

\bibitem{veerapaneni:2011aa}
Veerapaneni SK, Rahimian A, Biros G, Zorin D (2011) A fast algorithm for
  simulating vesicle flows in three dimensions.
\newblock Journal of Computational Physics 230: 5610 - 5634.
\input{veerapaneni:2011aa}

\bibitem{danker:2009ep}
Danker G, Vlahovska P (2009) Vesicles in poiseuille flow.
\newblock Physical review letters 102: 148102.
\input{danker:2009ep}

\bibitem{boedec2012settling}
Boedec G, Jaeger M, Leonetti M (2012) Settling of a vesicle in the limit of
  quasispherical shapes.
\newblock Journal of Fluid Mechanics 690: 227--261.
\input{boedec2012settling}

\bibitem{huang2011sedimentation}
Huang Z, Abkarian M, Viallat A (2011) Sedimentation of vesicles: from pear-like
  shapes to microtether extrusion.
\newblock New Journal of Physics 13: 035026.
\input{huang2011sedimentation}

\bibitem{kushner:2001pm}
Kushner J, Rother M, Davis R (2001) Buoyancy-driven interactions of viscous
  drops with deforming interfaces.
\newblock Journal of Fluid Mechanics 446: 253--269.
\input{kushner:2001pm}

\bibitem{rother:2004ll}
Rother M, Davis R (2004) Buoyancy-driven coalescence of spherical drops covered
  with incompressible surfactant at arbitrary p{\'e}clet number.
\newblock Journal of Colloids and Interface Science 270: 205--220.
\input{rother:2004ll}

\bibitem{rother:2006yk}
Rother MA, Zinchenko AZ, Davis RH (2006) Surfactant effects on buoyancy-driven
  viscous interactions of deformable drops.
\newblock Colloids and Surfaces A: Physicochemical and Engineering Aspects 282:
  50--60.
\input{rother:2006yk}

\bibitem{rother:2008yv}
Rother MA, Davis RH (2008) Buoyancy-driven breakup of an isolated drop with
  surfactant.
\newblock Physical Review Letters 101: 44501.
\input{rother:2008yv}

\bibitem{ascoli:1990xy}
Ascoli EP, Dandy DS, Leal LG (1990) Buoyancy-driven motion of a deformable drop
  toward a planar wall at low reynolds number.
\newblock Journal of Fluid Mechanics 213: 287--311.
\input{ascoli:1990xy}

\bibitem{pozrikidis:1992rz}
Pozrikidis C (1992) The buoyancy-driven motion of a train of viscous drops
  within a cylindrical tube.
\newblock Journal of Fluid Mechanics 237: 627--648.
\input{pozrikidis:1992rz}

\bibitem{zinchenko:1997lk}
Zinchenko A, Rother M, Davis R (1997) A novel boundary-integral algorithm for
  viscous interaction of deformable drops.
\newblock Physics of Fluids 9: 1493.
\input{zinchenko:1997lk}

\bibitem{zinchenko:1999pr}
Zinchenko AZ, Rother MA, Davis RH (1999) Cusping, capture, and breakup of
  interacting drops by a curvatureless boundary-integral algorithm.
\newblock Journal of Fluid Mechanics 391: 249--292.
\input{zinchenko:1999pr}

\bibitem{ly:2004fh}
Ly HV, Longo ML (2004) The influence of short-chain alcohols on interfacial
  tension, mechanical properties, area/molecule, and permeability of fluid
  lipid bilayers.
\newblock Biophysical Journal 87: 1013--1033.
\input{ly:2004fh}

\bibitem{evans:1990gl}
Evans E, Rawicz W (1990) Entropy-driven tension and bending elasticity in
  condensed-fluid membranes.
\newblock Physical Review Letters 64: 2094--2097.
\input{evans:1990gl}

\bibitem{zhong-can:1989zn}
Zhong-Can OY, Helfrich W (1989) Bending energy of vesicle membranes: General
  expressions for the first, second, and third variation of the shape energy
  and applications to spheres and cylinders.
\newblock Physical Review A 39: 5280--5288.
\input{zhong-can:1989zn}

\bibitem{derjaguin:1940mk}
Derjaguin B (1940) On the repulsive forces between charged colloid particles
  and on the theory of slow coagulation and stability of lyophobe sols.
\newblock Transactions of the Faraday Society 35: 203--215.
\input{derjaguin:1940mk}

\bibitem{carnie1994electrical}
Carnie S, Chan D, Gunning J (1994) Electrical double layer interaction between
  dissimilar spherical colloidal particles and between a sphere and a plate:
  The linearized poisson-boltzmann theory.
\newblock Langmuir 10: 2993--3009.
\input{carnie1994electrical}

\bibitem{stankovich1996electrical}
Stankovich J, Carnie S (1996) Electrical double layer interaction between
  dissimilar spherical colloidal particles and between a sphere and a plate:
  Nonlinear poisson-boltzmann theory.
\newblock Langmuir 12: 1453--1461.
\input{stankovich1996electrical}

\bibitem{menger1998giant}
Menger F, Angelova M (1998) Giant vesicles: imitating the cytological processes
  of cell membranes.
\newblock Accounts of Chemical Research 31: 789--797.
\input{menger1998giant}

\bibitem{greger2007basic}
Greger K, Swoger J, Stelzer E (2007) Basic building units and properties of a
  fluorescence single plane illumination microscope.
\newblock Review of Scientific Instruments 78: 023705.
\input{greger2007basic}

\bibitem{bradski2000opencv}
Bradski G (2000) The opencv library.
\newblock Doctor Dobbs Journal 25: 120--126.
\input{bradski2000opencv}

\bibitem{nelder1965simplex}
Nelder J, Mead R (1965) A simplex method for function minimization.
\newblock The Computer Journal 7: 308.
\input{nelder1965simplex}

\bibitem{pozrikidis:1992tc}
Pozrikidis C (1992) Boundary integral and singularity methods for linearized
  viscous flow.
\newblock New York: Cambridge University Press, first edition.
\input{pozrikidis:1992tc}

\bibitem{henriksen:2004jq}
Henriksen J, Rowat A, Ipsen J (2004) Vesicle fluctuation analysis of the
  effects of sterols on membrane bending rigidity.
\newblock European Biophysics Journal 33: 732--741.
\input{henriksen:2004jq}

\bibitem{olsen2009perturbations}
Olsen B, Schlesinger P, Baker N (2009) Perturbations of membrane structure by
  cholesterol and cholesterol derivatives are determined by sterol orientation.
\newblock Journal of the American Chemical Society 131: 4854--4865.
\input{olsen2009perturbations}

\bibitem{blake1971note}
Blake J (1971) A note on the image system for a stokeslet in a no-slip
  boundary.
\newblock In: Proc. Camb. Phil. Soc. Cambridge Univ Press, volume~70, pp.
  303--310.
\input{blake1971note}

\bibitem{seifert:1999sm}
Seifert U (1999) Fluid membranes in hydrodynamic flow fields: Formalism and an
  application to fluctuating quasispherical vesicles in shear flow.
\newblock The European Physical Journal B-Condensed Matter and Complex Systems
  8: 405--415.
\input{seifert:1999sm}

\bibitem{zhao:2011the-dynamics}
Zhao H, Shaqfeh E (2011) The dynamics of a vesicle in simple shear flow.
\newblock Journal of Fluid Mechanics 674: 578--604.
\input{zhao:2011the-dynamics}

\bibitem{cohen2003electrophoretic}
Cohen J (2003) Electrophoretic characterization of liposomes.
\newblock Methods in enzymology 367: 148--176.
\input{cohen2003electrophoretic}

\bibitem{iler1979chemistry}
Iler R (1979) The chemistry of silica: solubility, polymerization, colloid and
  surface properties, and biochemistry.
\newblock Wiley New York.
\input{iler1979chemistry}

\bibitem{polin2007colloidal}
Polin M, Grier D, Han Y (2007) Colloidal electrostatic interactions near a
  conducting surface.
\newblock Physical Review E 76: 041406.
\input{polin2007colloidal}

\bibitem{behrens2001charge}
Behrens S, Grier D (2001) The charge of glass and silica surfaces.
\newblock The Journal of Chemical Physics 115: 6716.
\input{behrens2001charge}

\bibitem{groves1998electric}
Groves J, Boxer S, McConnell H (1998) Electric field-induced critical demixing
  in lipid bilayer membranes.
\newblock Proceedings of the National Academy of Sciences 95: 935.
\input{groves1998electric}

\bibitem{trizac2002simple}
Trizac E, Bocquet L, Aubouy M (2002) Simple approach for charge renormalization
  in highly charged macroions.
\newblock Physical Review Letters 89: 248301.
\input{trizac2002simple}

\bibitem{tellez:2011aa}
Tellez G (2011) Nonlinear screening of charged macromolecules.
\newblock Philosophical Transactions of the Royal Society A: Mathematical,
  Physical and Engineering Sciences 369: 322--334.
\input{tellez:2011aa}

\bibitem{pozrikidis:2001hs}
Pozrikidis C (2001) Interfacial dynamics for stokes flow.
\newblock Journal of Computational Physics 169: 250--301.
\input{pozrikidis:2001hs}

\bibitem{meyer:2002vv}
Meyer M, Desbrun M, Schr{\"o}der P, Barr AH (2002) Discrete
  differential-geometry operators for triangulated 2-manifolds.
\newblock Visualization and mathematics 3: 34--57.
\input{meyer:2002vv}

\bibitem{abramoff:2004fk}
Abramoff M, Magalhaes P, Ram S (2004) Image processing with imagej.
\newblock Biophotonics International 11: 36--42.
\input{abramoff:2004fk}

\bibitem{rowat:2004fk}
Rowat AC, Hansen PL, Ipsen JH (2004) Experimental evidence of the electrostatic
  contribution to membrane bending rigidity.
\newblock EPL (Europhysics Letters) 67: 144.
\input{rowat:2004fk}

\bibitem{mitchell:1989aa}
Mitchell D, Ninham B (1989) Curvature elasticity of charged membranes.
\newblock Langmuir 5: 1121--1123.
\input{mitchell:1989aa}

\bibitem{winterhalter:1988wj}
Winterhalter M, Helfrich W (1988) Effect of surface charge on the curvature
  elasticity of membranes.
\newblock The Journal of Physical Chemistry 92: 6865--6867.
\input{winterhalter:1988wj}

\bibitem{pan2008cholesterol}
Pan J, Mills T, Tristram-Nagle S, Nagle J (2008) Cholesterol perturbs lipid
  bilayers nonuniversally.
\newblock Physical Review Letters 100: 198103.
\input{pan2008cholesterol}

\end{thebibliography}

\section*{Figure Legends}

\end{document}